\documentstyle[12pt]{article}
\newlength{\extraspace}
\newlength{\extraspaces}
\newcommand{\be}{\begin{equation}
\addtolength{\abovedisplayskip}{\extraspaces}
\addtolength{\belowdisplayskip}{\extraspaces}
\addtolength{\abovedisplayshortskip}{\extraspace}
\addtolength{\belowdisplayshortskip}{\extraspace}}
\newcommand{\ee}{\end{equation}}
\newcommand{\ba}{\begin{eqnarray}
\addtolength{\abovedisplayskip}{\extraspaces}
\addtolength{\belowdisplayskip}{\extraspaces}
\addtolength{\abovedisplayshortskip}{\extraspace}
\addtolength{\belowdisplayshortskip}{\extraspace}}
\newcommand{\ea}{\end{eqnarray}}
\newcommand{\nonu}{\nonumber \\[.5mm]}
\newcommand{\A}{&\!\!\!}

\begin{document}
\addtolength{\baselineskip}{.7mm}
\begin{flushright}
STUPP-97-152 \\ November, 1997
\end{flushright}
\vspace{.6cm}
\begin{center}
{\large{\bf{Minimal Off-Shell Version of 
            $N = 1$ Chiral Supergravity}}} \\[20mm]
{\sc Motomu Tsuda and Takeshi Shirafuji} \\[12mm]
{\it Physics Department, Saitama University \\[2mm]
Urawa, Saitama 338, Japan} \\[20mm]
{\bf Abstract}\\[10mm]
{\parbox{13cm}{\hspace{5mm}
We construct the minimal off-shell formulation of $N = 1$ 
chiral supergravity (SUGRA) introducing 
a complex antisymmetric tensor field $B_{\mu \nu}$ 
and a complex axial-vector field $A_{\mu}$ 
as auxiliary fields. The resulting algebra of the right- 
and left-handed supersymmetry (SUSY) transformations 
closes off shell and generates chiral gauge 
transforamtions and vector gauge transformations 
in addition to the transformations which appear 
in the case without auxiliary fields. }} 
\end{center}
\vfill

%%%%%%%%%%%%%%%%%%%%%%%%%%%%%%%%%%%%%%%%%%%%%%%%%%%%%%%%%%%%
\newpage
%\setcounter{section}{0}
%\setcounter{equation}{0}
%
%\newsection{}

$N = 1$ chiral supergravity (SUGRA), in which only a right-handed 
spin-3/2 field is coupled to gravitational field, 
was first formulated by Jacobson \cite{JJ} 
as the extension of Ashtekar's canonical formulation 
of general relativity \cite{AA,JS}. 
There appear two kinds of supersymmetry (SUSY) transformations, 
namely, right- and left-handed SUSY transformations. 
After that, the first-order formulation of $N = 1$ chiral SUGRA 
was made using the method of the two-form gravity \cite{CDJ,KS}. 
However, in the two-form SUGRA, the SUSY transformation 
parameters are constrained and the problem of the off-shell 
closure of SUSY algebra is still open. 

In a previous paper \cite{TS}, we constructed the first-order 
formulation of $N = 1$ chiral SUGRA following 
the usual $N = 1$ SUGRA \cite{FN,DZ}. 
In this formulation, the SUSY transformation parameters 
are not constrained at all in contrast with the method 
of the two-form gravity, although we introduce a complex tetrad 
and take a chiral Lagrangian as analytic in field variables. 
We also studied the SUSY algebra in the second-order formulation 
and showed that it closes only on shell 
as extra terms proportional to spin-3/2 field equations 
appear in the algebra. 
These results suggest that we need additional auxiliary 
fields in $N = 1$ chiral SUGRA as in the usual one. 

There exist two minimal off-shell versions in the usual 
$N = 1$ SUGRA, namely, one with the set of auxiliary fields 
$(S, P, A_{\mu})$ \cite{SWFN} and the other with the set 
of auxiliary fields $(B_{\mu \nu}, A_{\mu})$ \cite{SW}. 
We show in this letter that it is possible to construct 
a minimal off-shell version of $N = 1$ chiral SUGRA 
provided one uses a complex antisymmetric tensor field 
$B_{\mu \nu}$ and a complex axial-vector field $A_{\mu}$ 
as auxiliary fields. 
We also note that the set of auxiliary fields $(S, P, A_{\mu})$ 
does not suit for $N = 1$ chiral SUGRA. 

Let us begin with the linearized off-shell theory 
of $N = 1$ chiral SUGRA. 
The field variables are complex gravitational field 
$h_{\mu \nu}$, two independent (Majorana) Rarita-Schwinger 
fields $(\psi_{R \mu}, \tilde \psi_{L \mu})$, 
\footnote{\ We assume $\psi_{\mu}$ 
and $\tilde \psi_{\mu}$ to be two independent 
(Majorana) Rarita-Schwinger fields, 
and define the right-handed spinor fields 
$\psi_{R \mu} := (1/2)(1 + \gamma_5) \psi_{\mu}$ 
and $\tilde \psi_{R \mu}:= (1/2)(1 + \gamma_5) 
\tilde \psi_{\mu}$. 
The $\psi_{R \mu}$ and $\tilde \psi_{R \mu}$ 
are related to the left-handed spinor fields 
$\psi_{L \mu}$ and $\tilde \psi_{L \mu}$ respectively, 
because $\psi_{\mu}$ and $\tilde \psi_{\mu}$ 
are Majorana spinors. 
We shall follow the notation and convention 
of Ref.\cite{TS}.}
and the complex auxiliary fields $(B_{\mu \nu}, A_{\mu})$. 
The two kinds of linearized SUSY transformations can be 
defined as follows. The right-handed rigid SUSY 
transformations are generated by a constant Majorana 
spinor parameter $\alpha_R$ and take the form 
\ba
\A \A \delta_R h_{\mu \nu} 
      = {i \over 2} (\overline \alpha_L \gamma_{\mu} 
       \tilde \psi_{L \nu} + \overline \alpha_L \gamma_{\nu} 
      \tilde \psi_{L \mu}), \nonu
\A \A \delta_R \psi_{R \mu} 
      = -2i \ S^{\rho \sigma} \partial_{\rho} h_{\sigma \mu} 
      \alpha_R + i(A_{\mu} + V_{\mu}) \alpha_R 
      + S_{\mu \nu} V^{\nu} \alpha_R, \nonu
\A \A \delta_R \tilde \psi_{L \mu} = 0, \nonu
\A \A \delta_R A_{\mu} 
      = 2i \ \overline \alpha_L \gamma_{\mu} 
      S_{\rho \sigma} \partial^{\rho} 
      \tilde {\psi_L}{^\sigma}, \nonu
\A \A \delta_R B_{\mu \nu} 
      = 2i \ (\overline \alpha_L \gamma_{\mu} 
      \tilde \psi_{L \nu} - \overline \alpha_L \gamma_{\nu} 
      \tilde \psi_{L \mu}) 
\label{linR}
\ea
with $V^{\mu}$ being defined by 
\be
V^{\mu} := {1 \over 4} \epsilon^{\mu \nu \rho \sigma} 
           \partial_{\nu} B_{\rho \sigma}. 
\label{V}
\ee
On the other hand, the left-handed rigid SUSY transformations, 
which are generated by a constant Majorana spinor parameter 
$\tilde \alpha_L$, are given by 
\ba
\A \A \delta_L h_{\mu \nu} 
      = {i \over 2} (\overline{\tilde \alpha}_R \gamma_{\mu} 
      \psi_{R \nu} + \overline{\tilde \alpha}_R \gamma_{\nu} 
      \psi_{R \mu}), \nonu
\A \A \delta_L \psi_{R \mu} = 0, \nonu
\A \A \delta_L \tilde \psi_{L \mu} 
      = -2i \ S^{\rho \sigma} \partial_{\rho} h_{\sigma \mu} 
      \tilde \alpha_L - i(A_{\mu} + V_{\mu}) \tilde \alpha_L 
      - S_{\mu \nu} V^{\nu} \tilde \alpha_L, \nonu
\A \A \delta_L A_{\mu} 
      = - 2i \ \overline{\tilde \alpha}_R 
      \gamma_{\mu} S_{\rho \sigma} 
      \partial^{\rho} {\psi_R}^{\sigma}, \nonu
\A \A \delta_L B_{\mu \nu} 
      = 2i \ (\overline{\tilde \alpha}_R \gamma_{\mu} 
      \psi_{R \nu} - \overline{\tilde \alpha}_R \gamma_{\nu} 
      \psi_{R \mu}). 
\label{linL} 
\ea
The algebra of SUSY transformations (\ref{linR}) 
and (\ref{linL}) closes off shell, i.e., 
without using the field equations 
for $\psi_R$ and $\tilde \psi_L$. 

The linearized theory of $N = 1$ chiral SUGRA with 
the auxiliary fields $(B_{\mu \nu}, A_{\mu})$ are also 
invariant under gauge transformations 
\ba
\A \A \delta_g h_{\mu \nu} = \partial_{\mu} a_{\nu} 
      + \partial_{\nu} a_{\mu}, 
\label{gauge-1} \\
\A \A \delta_g \psi_{R \mu} = \partial_{\mu} \epsilon_R, \ \ 
      \delta_g \tilde \psi_{L \mu} 
      = \partial_{\mu} \tilde \epsilon_L, 
\label{gauge-2} \\
\A \A \delta_g A_{\mu} = \partial_{\mu} \Lambda, 
\label{gauge-3} \\
\A \A \delta_g B_{\mu \nu} = \partial_{\mu} b_{\nu} 
      - \partial_{\nu} b_{\mu} 
\label{gauge-4}
\ea
with five independent parameters $a_{\mu}, 
\epsilon_R, \tilde \epsilon_L, \Lambda$ and $b_{\mu}$. 

The linearized chiral Lagrangian, which is invariant 
under the right- and left-handed SUSY transformations 
and the gauge transformations, can be written as 
\be
L^{(+)} = - {1 \over 2} h^{\mu \nu} G_{\mu \nu} 
          - \epsilon^{\mu \nu \rho \sigma} 
          \overline{\tilde \psi}_{R \mu} \gamma_\rho 
          \partial_\sigma \psi_{R \nu} 
          + {3 \over 4} V_{\mu} V^{\mu} + A_{\mu} V^{\mu}, 
\label{lin-Lag+}
\ee
where $G_{\mu \nu}$ is the linearized Einstein tensor, 
\be
G_{\mu \nu} = - \{ \Box {\overline h_{(\mu \nu)}} 
   - \partial^{\rho} (\partial_{\mu} {\overline h_{(\nu \rho)}} 
   + \partial_{\nu} {\overline h_{(\mu \rho)}}) 
   + \eta_{\mu \nu} \partial^{\rho} \partial^{\sigma} 
   {\overline h_{(\rho \sigma)}} \} 
\ee
with 
\be
{\overline h_{(\mu \nu)}} := h_{(\mu \nu)} 
- {1 \over 2} \eta_{\mu \nu} h, 
\ \ h := \eta^{\mu \nu} h_{(\mu \nu)}, 
\ee
and the d'Alembertian operator being defined 
by $\Box := \partial^{\mu} \partial_{\mu}$. 

The field equations at the linearized level 
can be derived from (\ref{lin-Lag+}) as 
\ba
\A \A G_{\mu \nu} = 0, \nonu
\A \A R_R^{\mu} := \epsilon^{\mu \nu \rho \sigma} 
                   \gamma_{\nu} \partial_{\rho} 
                   \psi_{R \sigma} = 0, \nonu
\A \A \tilde R_L^{\mu} := \epsilon^{\mu \nu \rho \sigma} 
                   \gamma_{\nu} \partial_{\rho} 
                   \tilde \psi_{L \sigma} = 0, \nonu
\A \A V^{\mu} = 0, \nonu
\A \A \epsilon^{\mu \nu \rho \sigma} 
      (3 \partial_{\nu} V_{\mu} + 2 \partial_{\nu} A_{\mu}) 
      = 0. 
\label{lin-eq}
\ea
>From the last two equations of (\ref{lin-eq}), 
we can see that the gauge invariant part of $B_{\mu \nu}$, 
namely $V^{\mu}$ of (\ref{V}), and that of $A_{\mu}$, namely 
$f_{\mu \nu} (:= \partial_{\mu} A_{\nu} 
- \partial_{\nu} A_{\mu})$ are zero. 
Therefore the auxiliary fields $(B_{\mu \nu}, A_{\mu})$ 
do not have on-shell degrees of freedom, while they have 
six (complex) off-shell degrees of freedom. 
This number is precisely the mismatch number of the components 
of bosonic fields ($h_{\mu \nu}$) and fermionic 
fields ($\psi_{R \mu}, \tilde \psi_{L \mu}$). 
Note that if the gravitational field $h_{\mu \nu}$ 
and the auxiliary fields $(B_{\mu \nu}, A_{\mu})$ 
are real, and if $\tilde \psi_{\mu} = \psi_{\mu}$, then 
the linearized off-shell theory of $N = 1$ chiral SUGRA 
is reduced to that of the usual $N = 1$ SUGRA \cite{SW}. 

One method to construct the full nonlinear theory 
in the usual $N = 1$ SUGRA starts from the linearized theory 
and makes the rigid SUSY transformations local, 
adding appropriate terms to the linearized Lagrangian 
order by order in the gravitational constant $\kappa$ 
\cite{West}. 
\footnote{\ The $\kappa^2$ is the Einstein constant: 
$\kappa^2 = 8 \pi G/c^4$.}
We expect that the full nonlinear theory of $N = 1$ chiral SUGRA 
with $(B_{\mu \nu}, A_{\mu})$ can be obtained 
>from its linearized theory in the similar manner. 
However we can deduce the full nonlinear form 
of SUSY transformations directly from the linearized transformation 
laws of (\ref{linR}) and (\ref{linL}), 
referring to the corresponding result of the usual $N = 1$ SUGRA. 

In the full nonlinear theory, the field variables are 
the complex tetrad $e^i_{\mu}$, 
two (Majorana) Rarita-Schwinger fields $(\psi_{R \mu}, 
\tilde \psi_{L \mu})$ 
and the complex auxiliary fields $(B_{\mu \nu},$ $A_{\mu})$. 
We choose the right-handed SUSY transformations 
generated by $\alpha_R$ to be 
\ba
\A \A \delta_R e^i_{\mu} 
      = i \kappa \ \overline \alpha_L \gamma_i 
      \tilde \psi_{L \mu}, \nonu
\A \A \delta_R \psi_{R \mu} = {2 \over \kappa} 
      {\cal D}^{(+)}_{\mu} \alpha_R, 
      \ \ \ \ \ \delta_R \tilde \psi_{L \mu} = 0, \nonu
\A \A \delta_R A_{\mu} = -{i \over 2} \ \overline \alpha_L 
      \gamma_{\mu} \gamma \cdot \tilde {\cal R}_L, \nonu
\A \A \delta_R B_{\mu \nu} = 2i \ (\overline \alpha_L 
      \gamma_{\mu} \tilde \psi_{L \nu} 
      - \overline \alpha_L \gamma_{\nu} \tilde \psi_{L \mu}), 
\label{fullR}
\ea
and the left-handed SUSY transformations generated by 
$\tilde \alpha_L$ to be 
\ba
\A \A \delta_L e_{\mu}^i 
      = i \kappa \ \overline{\tilde \alpha}_R 
      \gamma^i \psi_{R \mu}, \nonu
\A \A \delta_L \psi_{R \mu} = 0, 
      \ \ \ \ \ \delta_L \tilde \psi_{L \mu} 
      = {2 \over \kappa} {\cal D}^{(-)}_{\mu} 
      \tilde \alpha_L, \nonu
\A \A \delta_L A_{\mu} = -{i \over 2} \ \overline{\tilde \alpha}_R 
      \gamma_{\mu} \gamma \cdot {\cal R}_R, \nonu
\A \A \delta_L B_{\mu \nu} = 2i \ (\overline{\tilde \alpha}_R 
      \gamma_{\mu} \psi_{R \nu} 
      - \overline{\tilde \alpha}_R \gamma_{\nu} \psi_{R \mu}), 
\label{fullL}
\ea
where we define 
\ba
\A \A {\cal D}^{(+)}_{\mu} := D^{(+)}_{\mu} + {\kappa \over 2}
      \{ \ i(A_{\mu} + V_{\mu}) + S_{\mu \nu} V^{\nu} \}, \nonu
\A \A {\cal D}^{(-)}_{\mu} := D^{(-)}_{\mu} - {\kappa \over 2}
      \{ \ i(A_{\mu} + V_{\mu}) + S_{\mu \nu} V^{\nu} \}, \nonu
\A \A {\cal R}_R^{\mu} := \epsilon^{\mu \nu \rho \sigma} 
      \gamma_{\nu} {\cal D}^{(+)}_{\rho} \psi_{R \sigma}, \nonu
\A \A \tilde {\cal R}_L^{\mu} := \epsilon^{\mu \nu \rho \sigma} 
      \gamma_{\nu} {\cal D}^{(-)}_{\rho} \tilde \psi_{L \sigma}, 
\ea
and the $V^{\mu}$ is now given by 
\be
V^{\mu} := \epsilon^{\mu \nu \rho \sigma} \left( {1 \over 4} 
           \partial_{\nu} B_{\rho \sigma} - {i \over 2} \kappa 
           \ \overline{\tilde \psi}_{R \nu} \gamma_{\rho} 
           \psi_{R \sigma} \right). 
\label{full-V}
\ee
$D^{(+)}_\mu$ denotes 
the covariant derivative with respect to 
the self-dual connection $A^{(+)}_{ij \mu}$ 
which satisfies 
$(1/2){\epsilon_{ij}} \! ^{kl} A^{(+)}_{kl \mu} 
= i A^{(+)}_{ij \mu}$: 
\be
D^{(+)}_\mu := \partial_\mu + {i \over 2} A^{(+)}_{ij \mu} S^{ij} 
\ee
with $S^{ij}$ being the Lorentz generator. 
Here the self-dual connection $A^{(+)}_{ij \mu}$ represents that 
in the second-order formulation, namely, 
\be
A^{(+)}_{ij \mu} = A^{(+)}_{ij \mu}(e) + K^{(+)}_{ij \mu}, 
\label{A+K}
\ee
where $A^{(+)}_{ij \mu}(e)$ is the self-dual part 
of the Ricci rotation coefficients $A_{ij \mu}(e)$, 
while $K^{(+)}_{ij \mu}$ is that of $K_{ij \mu}$ given by 
\footnote{\ 
The antisymmetrization of a tensor with respect to $i$ 
and $j$ is denoted by $A_{[i \mid \cdots \mid j]} 
:= (1/2)(A_{i \cdots j} - A_{j \cdots i})$.}
\be
K_{ij \mu} := {i \over 2} (e^{\rho}_i e^{\sigma}_j e_{\mu}^k 
             \ \overline{\tilde \psi}_{R [\rho} 
                     \gamma_{\mid k \mid} \psi_{R \sigma]} 
    + e^{\rho}_i 
             \ \overline{\tilde \psi}_{R [\rho} 
                     \gamma_{\mid j \mid} \psi_{R \mu]} 
    - e^{\rho}_j 
             \ \overline{\tilde \psi}_{R [\rho} 
                     \gamma_{\mid i \mid} \psi_{R \mu]}). 
\ee
On the other hand, $D^{(-)}_\mu$ stands for 
the covariant derivative with respect to 
the anti-self-dual part of $A_{ij \mu}(e) + K_{ij \mu}$. 

As for the gauge transformations of the auxiliary fields 
$(B_{\mu \nu}, A_{\mu})$ defined by (\ref{gauge-3}) and 
(\ref{gauge-4}) in the linearized off-shell theory, 
we take in the full nonlinear theory as follows: 
\be
\delta_V B_{\mu \nu} = \partial_{\mu} b_{\nu} 
- \partial_{\nu} b_{\mu} 
\ee
for vector gauge transformations and 
\ba
\A \A \delta_C A_{\mu} 
      = -{2 \over \kappa} \partial_{\mu} \Lambda, \nonu
\A \A \delta_C \psi_{R \mu} = i \Lambda \psi_{R \mu}, \ \ \ \ \ 
      \delta_C \tilde \psi_{L \mu} 
      = - i \Lambda \tilde \psi_{L \mu} 
\label{c-gauge}
\ea
for chiral gauge transformations, 
where $\psi_{R \mu}$ and $\tilde \psi_{L \mu}$ need to be 
transformed as in (\ref{c-gauge}) because the definition of $V^{\mu}$ 
is modified as (\ref{full-V}). 

It can now be shown that the commutator algebra of the right- 
and left-handed SUSY transformations (\ref{fullR}) 
and (\ref{fullL}) closes off shell as 
\ba
\A \A [\delta_{R(1)}, \delta_{R(2)}] = 0 
      = [\delta_{L(1)}, \delta_{L(2)}], \\
\A \A [\delta_{R(1)}, \delta_{L(2)}] = \delta_G (\xi_{\mu}) 
      + \delta_{{\rm Lorentz}} (\xi^{\rho} A_{ij \rho} 
      + {\kappa \over 2} \epsilon_{ijkl} \xi^k V^l) 
      + \delta_R (-{\kappa \over 2} \xi \cdot \psi_R) \nonu
\A \A \ \ \ \ \ \ \ \ \ \ \ \ \ \ \ \ \ \ 
      + \delta_L (-{\kappa \over 2} \xi \cdot \tilde \psi_L) 
      + \delta_C ({\kappa \over 2} \xi \cdot A) 
      + \delta_V (-{2 \over \kappa} \xi_{\mu} 
      + B_{\mu \nu} \xi^{\nu}) 
\label{com-RL}
\ea
with 
\be
\xi^i := 2i \ \overline{\tilde \alpha}_{2R} 
\gamma^i \alpha_{1R}. 
\ee
Note that $\xi^i$ is complex because $\alpha_{1R}$ and 
$\overline{\tilde \alpha}_{2R}$ are independent of each other. 
The commutator (\ref{com-RL}) generates the general 
coordinate transformations ($\delta_G$), the local 
Lorentz transformations ($\delta_{{\rm Lorentz}}$), 
the right- and left-handed SUSY transformations 
($\delta_R$, $\delta_L$), the chiral gauge transformations 
($\delta_C$) and the vector gauge transformations ($\delta_V$). 
The quantities in the parentheses in (\ref{com-RL}) 
denote the parameters of the respective transformations. 
The characteristic feature of the SUSY algebra 
in $N = 1$ chiral SUGRA is that the right- or left-handed 
SUSY transformations commute with each other, 
while the commutator of the right- and left-handed 
SUSY transformations generates all invariant transformations 
in the theory. 

The chiral Lagrangian density invariant under all the 
transformations generated by the commutator (\ref{com-RL}) 
has the following form: 
\be
{\cal L}^{(+)} = -{i \over {2 \kappa^2}} e \ 
                 \epsilon^{\mu \nu \rho \sigma} 
                 e_{\mu}^i e_{\nu}^j R^{(+)}_{ij \rho \sigma} 
                 - e \ \epsilon^{\mu \nu \rho \sigma} 
                 \overline{\tilde \psi}_{R \mu} \gamma_\rho 
                 D^{(+)}_\sigma \psi_{R \nu} 
                 + {3 \over 4} e V_{\mu} V^{\mu} + e A_{\mu} V^{\mu}, 
\label{full-Lag+}
\ee
where $e$ denotes ${\rm det}(e^i_{\mu})$ 
and the curvature of the self-dual connection 
${R^{(+)ij}}_{\mu \nu}$ is 
\be
{R^{(+)ij}}_{\mu \nu} := 2(\partial_{[\mu} {A^{(+)ij}}_{\nu]} 
             + {A^{(+)i}}_{k [\mu} {A^{(+)kj}}_{\nu]}). 
\ee
The field equation for $A^{(+)}_{ij \mu}$ gives (\ref{A+K}) 
and is used when we prove the invariance of the chiral Lagrangian 
density under the right- and left-handed SUSY transformations. 
It is noted that the linearized limit of (\ref{full-Lag+}) 
just becomes the linearized chiral Lagrangian of 
(\ref{lin-Lag+}). 

Brief remarks are in order about the field equations 
derived from the chiral Lagrangian density (\ref{full-Lag+}). 
The field equations for the auxiliary fields 
$(B_{\mu \nu}, A_{\mu})$ are given by 
\ba
\A \A V^{\mu} = 0 \ \ \ \ \ \ \ \ \ \ \ \ \ \ \ 
\ \ \ \ \ \ \ \ \ \ \ \ \ \ \ 
{\rm for} \ \ A_{\mu}, 
\label{A-eq} \\
\A \A e \epsilon^{\mu \nu \rho \sigma} 
      (3 \partial_{\nu} V_{\mu} - f_{\mu \nu}) = 0 
      \ \ \ \ \ \ \ \ {\rm for} \ \ B_{\mu \nu} 
\label{B-eq}
\ea
with $f_{\mu \nu} := \partial_{\mu} A_{\nu} 
- \partial_{\nu} A_{\mu}$, which shows that the gauge invariant 
part of $A_{\mu}$, namely $f_{\mu \nu}$, is zero, 
and that the gauge invariant part of $B_{\mu \nu}$ can be 
expressed in the bilinear form of spin-3/2 fields, 
i.e., $\overline{\tilde \psi}_{R \nu} 
\gamma_{\rho} \psi_{R \sigma}$. 
Secondly the field equations for all the field variables 
are reduced to those of the usual $N = 1$ SUGRA 
with $(B_{\mu \nu}, A_{\mu})$ 
provided that the following conditions are satisfied: 

(a) The tetrad $e^i_{\mu}$ and the auxiliary 
fields $(B_{\mu \nu}, A_{\mu})$ are real. 

(b) $\tilde \psi_{\mu} = \psi_{\mu}$. 

(c) The self-dual connection $A^{(+)}_{ij \mu}$ satisfies 
its equation of motion.

\newpage

We have constructed the minimal off-shell formulation 
of $N = 1$ chiral SUGRA by using the complex auxiliary fields 
$(B_{\mu \nu}, A_{\mu})$. However we cannot use 
the set of $(S, P, A_{\mu})$ as auxiliary fields 
in $N = 1$ chiral SUGRA. 
In order to see this, we remind of the extra terms which are 
proportional to spin-3/2 field equations in the SUSY algebra 
without auxiliary fields. If the set of $(S, P, A_{\mu})$ 
is used as auxiliary fields, the right-handed SUSY 
transformations of $S$ and $P$, for example, involve 
the term of $\overline \alpha_L \gamma \cdot R_R 
(R_R^{\mu} := \epsilon^{\mu \nu \rho \sigma} 
\gamma_{\nu} D^{(+)}_{\rho} \psi_{R \sigma})$. 
Since $\overline \alpha_L$ is paired with $\psi_{R \mu}$, 
this form cannot produce the appropriate counter terms 
to cancel the extra terms in the SUSY algebra. 
Therefore the $(S, P, A_{\mu})$ does not work 
as auxiliary fields. 

Finally we comment on future problems in chiral SUGRA. 
In the usual $N = 1$ SUGRA, there is the conformal 
theory \cite{KTN} and its close relationship 
to the various $N = 1$ Poincar\'e SUGRA 
with auxiliary fields has been clarified \cite{KU}. 
Similarly if there exists a conformal version 
of $N = 1$ chiral SUGRA, then the structure of 
chiral SUGRA will become clear and the matter-coupling 
problem will be simplified. 
The construction of such a conformal theory 
of $N = 1$ chiral SUGRA is now being investigated. 

The second problem is the formulation of extended chiral SUGRA. 
As for $N = 2$ case, the linearized theory can be 
constructed by means of the SUSY transformation 
parameters without constraints. 
We are now trying to construct the full nonlinear theory.

We would like to thank Professor Y. Tanii and other members 
of Physics Department at Saitama University 
for discussions and encouragements.

%%%%%%%%%%%%%%%%%%%%%%%%%%%%%%%%%%%%%%%%%%%%%%%%%%%%%%%%%%%%%%%%%%%%%
%%%%%%%%%%%%%%%%%%%%%%%%%%%%%%%%%%%%%%%%%%%%%%%%%%%%%%%%%%%%%%%%%%%%%

%%%%%%%%%%%%%%%%%%%%%%%%%%%%%%%%%%%%%%%%%%%%%%%%%%%%%%%%%%%%%%%%%%%%%
%%%%%%%%%%%%%%%%%%%%%%%%%%%%%%%%%%%%%%%%%%%%%%%%%%%%%%%%%%%%%%%%%%%%%


\newpage
\begin{thebibliography}{100}

%\bibitem{AAN} 
%T. Jacobson, in New Perspectives in Canonical Gravity, 
%edited by A. Ashtekar (Bibliopolis, Naples, 1988), p. 195. 

\bibitem{JJ} 
T. Jacobson, Class. Quantum Grav. 5 (1988) 923. 

\bibitem{AA} 
A. Ashtekar, Phys. Rev. Lett. 57 (1986) 2244; 
Phys. Rev. D 36 (1987) 1587. 

\bibitem{JS} 
T. Jacobson and L. Smolin, Phys. Lett. B 196 (1987) 39; 
Class. Quantum Grav. 5 (1988) 583. 

\bibitem{CDJ} 
R. Capovilla, J. Dell, T. Jacobson and L. Mason, 
Class. Quantum Grav. 8 (1991) 41. 

\bibitem{KS} 
H. Kunitomo and T. Sano, 
Int. J. Mod. Phys. D 1 (1993) 559. 

\bibitem{TS} 
M. Tsuda and T. Shirafuji, 
Phys. Rev. D 54 (1996) 2960. 

\bibitem{FN} 
D. Z. Freedman , P. van Nieuwenhuizen and S. Ferrara, 
Phys. Rev. D 13 (1976) 3214; 
D. Z. Freedman and P. van Nieuwenhuizen, 
Phys. Rev. D 14 (1976) 912. 

\bibitem{DZ} 
S. Deser and B. Zumino, Phys. Lett. B 62 (1976) 335. 

%\bibitem{Nieu} 
%P. van Nieuwenhuizen, Phys. Rep. C 68 (1981) 189. 

\bibitem{SWFN} 
K. S. Stelle and P. C. West, 
Phys. Lett. B 74 (1978) 330; 
S. Ferrara and P. van Nieuwenhuizen, 
Phys. Lett. B 74 (1978) 333. 

\bibitem{SW} 
M. F. Sohnius and P. C. West, 
Phys. Lett. B 105 (1981) 353. 

\bibitem{West} 
P. West, 
Introduction to Supersymmetry and Supergravity 
(World Scientific, Singapore, 1990). 

\bibitem{KTN}
M. Kaku, P. K. Townsend and P. van Nieuwenhuizen, 
Phys. Rev. D 17 (1978) 3179. 

\bibitem{KU}
T. Kugo and S. Uehara, 
Nucl. Phys. B 226 (1983) 49. 






\end{thebibliography}
\end{document}